\documentclass[a4paper,11pt]{article}
\usepackage{graphicx}

\usepackage{graphicx}

\def\bea{\begin{eqnarray}}
\def\eea{\end{eqnarray}}
\def\ket#1{\vert #1 \rangle}
\def\bra#1{\langle #1 \vert}

\def\sqr#1#2{{\vcenter{\vbox{\hrule height.#2pt
      \hbox{\vrule width.#2pt height#1pt \kern#1pt
         \vrule width.#2pt}
      \hrule height.#2pt}}}}

\def\figloc#1#2 {
\begin{figure}[!htb]\begin{center}
    \includegraphics[width=80mm]{fig#1.pdf}
    \caption{ #2}
    \end{center}\end{figure}
}

\begin{document}

\title{Quantum Noise in Amplifiers and Hawking/Dumb-Hole Radiation as Amplifier Noise}
\maketitle

{\author{W. G. Unruh\\
{CIAR Cosmology and Gravity Program\\
Dept.~of Physics and Astronomy\\
University of B. C.\\
Vancouver, Canada V6T 1Z1\\
unruh@physics.ubc.ca
\\
}}}

\textbf{Abstract:}
The quantum noise in a linear amplifier is shown to be thermal noise. The
theory of linear amplifiers is applied first to the simplest, single or double
oscillator model of an amplifier, and then to a linear model of an amplifier
with continuous fields and input and outputs. Finally it is shown that the
thermal 
noise emitted by black holes first demonstrated by Hawking, and of dumb holes
(sonic and other analogs to black holes), arises from the same analysis as for
linear amplifiers. The amplifier noise of black holes acting as amplifiers on the quantum
fields living in the spacetime surrounding the black hole   is the radiation
discovered by Hawking. For any amplifier, that quantum noise is completely
characterized by the attributes of the system regarded as a classical
amplifier, and arises out of those classical amplification factors and the
commutation relations of quantum mechanics. 

\section{Introduction}
Linear amplifiers, devices which take in a signal and produce and output signal of a
different amplitude,  are ubiquitous, but in general seem to be poorly
understood. All produce noise, but again the source of that noise tends to be
poorly understood, and seems often based on a case by case analysis. While
often an amplifier can have excess noise, caused by some infelicity in its
construction, all amplifiers must have a minimum level of noise, set by
quantum mechanics. This was recognized by Haus and
Mullen~\cite{haus} and others~\cite{caves,review,others,yamamoto} 
half a century ago,
but the lesson bears repeating. In particular, the noise is often
roughly characterized by a temperature. We will see that this is exact -- the
noise output is thermal. Furthermore, black holes turn out to simply be an
unusual instance of one of these amplifiers.

By a linear amplifier I mean a device into which one feeds an input signal
$I(t)$ and out of which comes an  amplified output signal $\int A(t-t') I(t')dt'$. In
principle, that amplification $A$ could be a function of both $t$ and $t'$
independently, 
rather than just their difference.  Such an amplifier is in general a phase
sensitive amplifier, for example, one which amplifies the cosine component of
the input differently from the sine component.  I will in this paper be interested phase
insensitive amplifiers as defined above.

All amplifiers are physically realised. Both the input and the output
signals are embodied by some physical quantities -- currents, voltages, light
intensities, magnetic fields, etc. One has some physical device into which one
places a physical signal and out of which comes an amplified signal. 
As the simplest model of an amplifier let me first assume that the input and
the output are both single modes, physical quantities defined by some single
degree of freedom which I will assume to be continuous (ie have a continuum of
possible values). I will assume that it is embodied by some quantum canonical
degree of freedom, designated by the operator $x$ which has a continuous
spectrum. Such a quantum degree of freedom will have a conjugate momentum
$p_x$ associated with it. Similarly the output will be assumed to be some
physical quantity which I will designate by 
the quantum  operator $Y$, with conjugate momentum $P_Y$. For example, $x$
might represent the charge on the capacitor in an LC circuit, and $P_x$ would
then be $LI= L {dx\over dt}$ where I is the current through the inductor.

 At present I will not be concerned with the time dependence of these
quantities, only that they obey a commutation relation, $[x,P_x]= i\hbar=i$
since I will take units such that $\hbar=1$.
The only ``signal" is the value
of the variable $x$. Now one operates on this system by some arbitrary
Hamiltonian, whose only requirement is that after the action of the amplifier,
there are a set of output dynamic variables $Y$, $P_Y$ which are related to the
input variables by $Y=Ax$,~~$P_Y=AP_x$. Both the dynamic variable $x$ and its
conjugate momentum are amplified after the interaction.
No matter what the input, the output is $A$ times larger.  Note that we are
working in the Heisenberg representation in which it is the dynamic variables
that change during an interaction, rather than the state that changes. 

The output variables could represent the same physical quantity, for example
the charge and L times the  current in the same LC circuit, just that the
interaction of the amplifier with that circuit has increased both the current
and the charge by the same amount. While this seems straightforward,  it is
immediately clear that this amplifier is unrealizable. Any amplifier is some physical
device which produces a unitary transformation between the input and the
output. In particular, if $Y$ and $P_Y$ are conjugate variables, we  have 
\bea
[Y,P_Y] =A^2[x,P_x]=A^2 i\hbar \,.
\eea
But clearly this is the correct commutation relation only if $A^2=1$, and the
amplification is trivial.

Does this mean that amplifiers cannot exist, in contradiction with experience?
The answer is of course ``No".  In order to have such
an amplifier one cannot have only one input channel. One needs at least two. 
(In this case a ``channel" just means another degree of freedom.) Consider
\bea
Y=Ax+Bq\,, \\
P_Y=AP_x+EP_q \,,
\eea
where $P_q$ is the conjugate momentum to $q$. Demanding that $[Y,P_Y]=i\hbar$
then leads to 
$$A^2+BE=1 \,.$$
We can always do a canonical transformation of the form $q\rightarrow e^\eta q$ and
$P_q\rightarrow e^{-\eta}P_q$, where $e^{2\eta}= \left\vert{E\over
B}\right\vert$. After this transformation, we have
\bea
E=-B\,,
\eea
so that
\bea
A^2-B^2=1\,,
\eea
or $A=\cosh(\mu)$ and $B=\sinh(\mu)$.

It is important here that the transformation by the amplifier of the
$q$,~$P_q$ channel to the $Y$, $P_Y$ channel be antilinear. Both are amplified
but the phase is reversed. 

Clearly if one has two input channels, one also needs two output channels as
well. Let me designate the second output channel by $Z, P_Z$. 
Then, in order that the commutation relations of $Z$ and $P_Z$ and $Y$ and
$P_Y$ be maintained, we can choose 
\bea
Z= Aq+Bx\,, \\
\tilde P_Z= AP_q-BP_x \,.
\eea
(One could also have other canonical transformations of $Z,\tilde P_Z$ which
would of course leave the commutation relations the same, but the one chosen is the
simplest case, and gives the same result as the others do).

This is more easily expressed in terms of creation and annihilation operators.
Defining 
\bea
a&=&{x+iP_x\over\sqrt{2}}\,,\\
b&=&{q+iP_q\over\sqrt{2}}\,,\\
C&=&{Y+iP_Y\over\sqrt{2}}\,,\\
D&=&{Z+iP_Z\over\sqrt{2}}\,,
\eea
 we have
\bea
C&=&\cosh(\mu) \;a +\sinh(\mu)\;b^\dagger\,,\\
D&=& \cosh(\mu)\;b +\sinh(\mu)\; a^\dagger\,,
\eea
or
\bea
a&=& \cosh(\mu)\;C -\sinh(\mu)\;D^\dagger\,, \\
b&=&\cosh(\mu)\;D -\sinh(\mu)\;C^\dagger\,.
\eea
This is just the form of a Bogoliubov transformation. 

Note that the commutation relations are maintained if we multiply $a$ or $b$
by phase
\bea
C&=&\cosh(\mu) \;e^{i\nu} a + \sinh(\mu) \;e^{-i\kappa} b^\dagger\,, \\
D&=& \cosh(\mu)\;e^{i\kappa}b +\sinh(\mu)\;e^{-i\nu} a^\dagger\,,
\eea
or multiply each term overall by a phase factor. In the following I will not
follow this complication, since all it does is to make the equations messier.

Let us now assume that the input states of the two input modes, represented by
$a, b$ are thermal states, with input density matrices 
\bea
\rho_a&=& N_a \;e^{-\Lambda_aa^\dagger a}\,, \\
\rho_b&=&N_b \;e^{-\Lambda_b b^\dagger b}\,.
\eea
 $N_a$ and $N_b$ are normalisation factors  equal to 
\bea
N_a= 1-e^{-\Lambda_a}\,, \\
N_b=1-e^{-\Lambda_b}\,,
\eea
so that ${\rm Tr}(\rho_a)={\rm Tr}(\rho_b)=1$.

If the dynamic variable corresponding to $x$ and thus  $a$ has a simple
Harmonic oscillator Hamiltonian with  frequency $\omega_a$, then we can write 
 $\Lambda_a={\omega_a / T_a}\,$,
where $T_a$ is the temperature of the $x$ input channel. Note that for most of
the following we do not have to make any assumptions about the Hamiltonian,
 since the results will only depend on $\Lambda$ and not on $T$.

The density matrix in terms of the output annihilation and creation operators
$C,~D$  is then
\bea
\rho&=&N_a\,N_b \,e^{-\Lambda_a a^\dagger a}e^{-\Lambda_b b^\dagger b}\\
&=&N_a\,N_b \,e^{-\Lambda_a a^\dagger a-\Lambda_b b^\dagger b}\\
&=&N_a\,N_b \, \exp\left[-\Lambda_a
\left(\cosh(\mu)C^\dagger-\sinh(\mu)D\right)\left(\cosh(\mu)C-\sinh(\mu)D^\dagger\right) \right. \nonumber \\ 
 && \left. -\Lambda_b\left(
\cosh(\mu)D^\dagger-\sinh(\mu)C\right)\left(\cosh(\mu)D-\sinh(\mu)C^\dagger\right)\right]\,.
\eea

Taking the trace over the  $D$ using the complete set of  $m$ particle states
defined by  $D^\dagger
D\ket{m}=m\ket{m}$, we have ${\rm Tr}_D(\rho)= \sum_m \bra{m}\rho\ket{m}$. Expanding
the exponential in $\rho$ 
in a power series, and expanding each term  so that each resultant term is of the form of a simple
product of the operators $C,~D,~C^\dagger,~D^\dagger$, we
see that each term must have the same number of $D$ as $D^\dagger$ operators.
Each $D$ lowers $m$ by 1, while each $D^\dagger$ increases $m$ by 1, and since 
($\bra{m}D^{\dagger r} D^s\ket{m}=0$ unless $r=s$, we need the same number of
$D$ and $D^\dagger$.
Since each $D$ in the expansion is multiplied by either a $D^\dagger$ or a $C$
and each  $D^\dagger$ by $D$ or $C^\dagger$, one must thus
also have the same number of $C$ as $C^\dagger$ operators in each term. After
commuting the $C$ and $C^\dagger$ appropriately so that the final expression
is a function of $C^\dagger C$, we see that the reduced density
matrix is thus a function only of $C^\dagger C$ (after an appropriate number
of commutations). It is in fact an exponential function in $C^\dagger C$  of
the form $e^{-\Lambda_C C^\dagger C}$ (see
appendix), where we can find $\Lambda_C$ from 
\bea
{\rm Tr} \left[C^\dagger \,C \, N_C\, e^{-\Lambda_C C^\dagger C}\right]&=&- N_C\,\partial_{\Lambda_C}
{\rm Tr}(e^{-\Lambda_C C^\dagger C})
\\
&=& {1\over e^{\Lambda_C}-1}\,.
\eea
Thus
\bea
{1\over e^{\Lambda_C}-1} &=& {\rm Tr}\left[N_aN_b \left\{ \cosh(\mu)^2 a^\dagger a
+\sinh(\mu)^2(b^\dagger b+1) + \right.\right. \nonumber \\ 
&& \left.\left. \cosh(\mu)\sinh(\mu) (ab+a^\dagger b^\dagger) \right\}
e^{-\Lambda_a a^\dagger a-\Lambda_bb^\dagger b}\right]\\
&=&{\cosh(\mu)^2 \over e^{\Lambda_a}-1 }+ \sinh(\mu)^2\left({1\over
e^{\Lambda_b}-1}+1\right)\,,
\eea
since the $a,b$ modes are uncorrelated. 

This is the crucial equation relating the output density matrix thermal factor  to the input
density matrices thermal factors. 
For small $\Lambda_{a,b}$ we get
\bea
{1\over \Lambda_C}= {\cosh(\mu)^2\over \Lambda_a} +\sinh(\mu)^2\left({1\over
\Lambda_b} +1\right)\,.
\eea

If the inputs and outputs all have the same frequency $\omega$, then
$\Lambda={\omega / T}$ and this becomes
\bea
T_C= \cosh(\mu)^2T_a + \sinh(\mu)^2 (T_b+\omega)\,,
\eea
and if the amplification is large, so that $\cosh(\mu)\approx\sinh(\mu)\approx
{e^{\mu}\over 2}$, 
\bea
{1\over\Lambda_C}={e^{2\mu}\over 4}\left({1\over \Lambda_a}+{1\over\Lambda_b}\right)\,,
\eea
or
\bea
T_C=\cosh(\mu)^2 \; (T_a+T_b)
\eea
(recall that $\cosh(\mu)$ is the amplification factor $A$).

For $\Lambda_{a,b}$ large (which corresponds to low temperatures) , we have
\bea
{1\over e^{\Lambda_C}-1}= \cosh(\mu)^2 e^{-\Lambda_a}+\sinh(\mu)^2\approx
\sinh(\mu)^2 \,,
\eea
or
\bea
\Lambda_C = -2\ln(\tanh(\mu)) \,.
\eea
This is exact in the limit as the input temperatures go to zero
($\Lambda_{a,b}\rightarrow\infty$).
Thus, for low temperatures in the inputs (temperatures much less than the input
frequencies), the output temperature is determined
by the amplification and the frequency of the output solely. This is what is
usually called  quantum
noise. Assuming the output has frequency $\omega_C$ we have
\bea
T_C={ \omega_C\over -2\ln(\tanh(\mu))} \,.
\eea
The output temperature of the $Y$ channel is completely determined by the
the amplification $\tanh(\mu)$. (Recall that $A=\cosh(\mu)$ was the
naive amplification of the amplifier.)

Note that the output thermal noise due to the amplification of the vacuum
fluctuation is given purely by the amplification $\cosh(\mu)$ and the
frequency of the output, both of which are determined purely by the classical
behaviour of the system and of the amplifier. The quantum noise of the amplifier is a
``classical" effect, in that it depends only on the classical attributes of the
amplifier. That the expression for the temperature includes a factor
 $\hbar\over k_B$ (suppressed in the above because of the choice of units)  does not alter the fact that it is completely determined
by classical attributes.

We can  put the input into a coherent state,
$a\ket{\alpha}=\alpha\ket{\alpha},\quad b\ket{\alpha}=0$. Then we have 
\bea
\bra{\alpha} C\ket{\alpha}=\cosh(\mu)\alpha\,,
\eea
where $\alpha$ can be as large as desired. That is, by measuring the output for a
classical input, one can determine the parameter of the amplifier which
determines the noise output of the amplifier. 

Alternatively one could have a situation in which  one takes
 $D$ as the output channel to be measured with $a$ still being the input
channel. Then the amplification of $a$ in the $D$ output is 
\bea
\bra{\alpha} D\ket{\alpha}=\sinh(\mu)\alpha \,.
\eea
That is, for small $\mu$ the ``amplification" goes to zero, rather than to 1.

It is also of interest to note that while the two inputs are, by assumption,
statistically independent (no correlations between $a,b$) the outputs are not. 
Even in the case of  vacuum input, we have
\bea
\left<CD\right> = \bra{0} \cosh(\mu)\sinh(\mu) (a a^\dagger + b^\dagger
b)\ket{0} =\cosh(\mu)\sinh(\mu) \,,
\eea
which implies a correlation (entanglement) between the $ C $ and $D$ outputs.

That same entanglement implies that we could have ``noiseless" amplification (ie, not
altering the signal to noise ratio of the input signal) by choosing an input
state which was an entangled state -- ie such that in the output state, the $C$
and $D$ modes were in a product state~\cite{squeezedinputs}. One would then
have a noiseless (zero temperature) output.  This is in general
not possible. In most physical systems the two input signals simply cannot be
correlated with each other. 
 However in certain situations, in which the signal is a classical
signal imposed on a quantum input channel, this may allow one to reduce the
noise in a detector by choosing an appropriately correlated set of input
channels, as for example in an interferometric gravity wave detection in
which the gravity wave signal, a very large ``classical" source, affects a
quantum input channel in the electromagnetic field in the arms of an
interferometer.

\section{Continuum}

While the above ``two mode" analysis is important, it is also instructive
to examine a model for a continuous phase insensitive amplifier -- ie, one with
a continuous, time dependent,  input signal which the amplifier continuously
amplifies into an output channel, as described in the first paragraphs.   I will present a
simple model for such an amplifier. The amplifier will be a
single-degree-of-freedom harmonic oscillator which couples, at $x=0$, two
massless scalar fields. While one of the fields is a normal scalar field, the
other will be one with negative energy. Its Lagrangian will be minus one
times the usual scalar field Lagrangian. It will act as the source of the
energy for the amplifier. Thus the two input channels will be fluctuations in
these fields travelling toward the oscillator, while the outputs will be the
same modes travelling away from the oscillator. 
 These propagating
modes  could, for example,  represent   electromagnetic
fields in a waveguide or 
 light
traveling toward a laser amplifier.

I will assume that the interaction between the oscillator and the two fields
is time independent. Thus in order to conserve energy while still amplifying
the signal, the 
the two fields must  have opposite signs of the
energy. Since an amplifier often feeds energy into the output mode -- that
energy must either come from the amplifier, or as here, come from the other
input channel.  

Thus, the Lagrangian for this amplifier model is
\bea
L&=&{1\over 2} \int \left[ \dot\phi^2-(\partial_x\phi)^2 -(\dot\psi^2
-(\partial_x\psi)^2) + 2\dot q (\epsilon\phi
+\tilde\epsilon\psi)\delta(x-\lambda) \right]dx \nonumber\\
&&\hskip .5cm +{1\over 2} (\dot q^2+\omega^2q^2) \,,
\eea
with reflection boundary condition at $x=0$ of  $\partial_x\psi(t,0)=\partial_x\phi(t,0)=0$. 
Here $\lambda$ is assumed to be very small, and we will take the limit as
$\lambda\rightarrow 0$.

I could have taken $x$ to be a continuous variable with the field propagating
from $-\infty$ to $+\infty$ and the oscillator located at $x=0$, but in that
case all of the antisymmetric modes for the $\phi$ and $\psi$ fields would not
have interacted at all with the oscillator.  In order to make sure that I have
the right boundary conditions at $x=0$, I take the oscillator to be located at
a small distance away from 0, namely $\lambda$,
and take the limit as $\lambda$ goes to zero.
The $\delta(x)$  is not well defined on the
half line $x\geq 0$. 

The $\psi$ field has negative definite energy, which is the source for the
energy amplification which accompanies the amplifier. (Note that while this
particular amplifier model amplifies the energy of the input, as well as its
amplitude, that is not necessary for an amplifier, as
we will see below.)

The equations of motion for the field are 
\bea
\partial_t^2\phi-\partial_x^2\phi &=&\epsilon\;\partial_t q\delta(x-\lambda) \,, \\
\partial_t^2\psi-\partial_x^2\psi &=&-\tilde\epsilon\;\partial_t
q\delta(x-\lambda) \,, \\
\partial_t^2 q +\Omega^2 q &=&-\epsilon\;\partial_t\phi(t,\lambda)
-\tilde \epsilon\;\partial_t\psi(t,\lambda) \,,
\eea
which have solutions   
\bea
\phi(t,x) &=& \phi_0(t,x)+\phi_0(t,-x) + \nonumber\\ && {1\over 2}\epsilon\left\{
\begin{array}{lr}
q(t-x-\lambda)+q(t-x+\lambda);& x>\lambda \\
 q(t+x-\lambda) +q(t-x-\lambda);& x<\lambda
 \end{array} \right\}\,,
\\
\psi(t,x) &=& \psi_0(t-x)+\psi_0(t+x) - \nonumber\\ && {1\over 2}\tilde\epsilon\left\{
\begin{array}{lr}
q(t-x-\lambda)+q(t-x+\lambda);& x>\lambda\\
q(t+x-\lambda) +q(t-x-\lambda);& x<\lambda
\end{array}\right\}\,,
\eea
and
\bea
\partial_t^2 q+\Omega^2 q&+&{\epsilon^2-\tilde\epsilon^2\over 2} \; \partial_t
(q(t)+q(t-2\lambda))=
-\epsilon \;\partial_t[\phi_0(t-\lambda)+ \nonumber\\ &&  \phi_0(t+\lambda)]
-\tilde\epsilon\; \partial_t[\psi_0(t-\lambda)+\psi_0(t+\lambda)]\,,
\eea
where I will only be interested in the limit as $\lambda\rightarrow 0$.
Taking the Fourier transform of the resulting equations where $q(t)=\int
q_\omega e^{-i\omega t}$, we have
\bea
q_\omega= 2i\omega{(\epsilon\phi_0(\omega) +\tilde\epsilon \psi_0(\omega))\over
-\omega^2-i\omega(\epsilon^2-\tilde\epsilon^2) +\Omega^2 } \,.
\eea

We take $\phi_0(t+x)$ and $\psi_0(t+x)$ as the ingoing modes (the $x$ and $q$ of the
above simple two mode analysis) and 
\bea
\phi_{out}= \psi_0(t-x)+\epsilon q(t-x)\,, \\
\psi_{out}=\psi_0(t-x) -\tilde \epsilon q(t-x) \,,
\eea
 are the outgoing modes corresponding to $C$, $D$ of the simple two mode
analysis.

Thus we find
\bea
\phi_{out}(\omega) = \psi_0(\omega) + 
2i\omega \epsilon{ \epsilon\phi_0(\omega) +\tilde\epsilon\psi_0(\omega) \over
-\omega^2-i(\epsilon^2-\tilde\epsilon^2)\omega+\Omega^2} \,.
\eea

The conserved norm for the system is
\bea
\left< \Xi',\Xi\right> &=&i\left[\int(\phi'^*\partial_t\phi -\psi'^*\partial_t\psi)dx
+q'^*(\partial_t q +\epsilon\phi(t,0) +\tilde\epsilon\psi(t,0)) \right. \\
&& - \left. \left(\int \partial_t\phi'^* \phi- \partial_t \psi'^*
\psi )dx  +(\partial_t
q'^* +\epsilon\phi'^*(t,0) +\tilde\epsilon\psi'^*(t,0))q\right) \right]\,, \nonumber
\eea
where $\Xi=\{\phi,\psi,q\}$ designates a complete solution of the equations of
motion at any time $t$. This norm is conserved by the equations of motion and
relates the ingoing modes at $t\rightarrow -\infty$ to the outgoing at
$t\rightarrow +\infty$.

Note the sign of the $\psi$ term in the norm. This arises from the fact that
the conjugate momentum for the $\psi$ field is $-\partial_t\psi$. 

The quantization of the the fields is such that positive norm fields are
associated with annihilation operators while the negative norm fields are
associated with creation operators. 
In the case of the $\psi$ field, the vacuum state, annihilated by the
annihilation operators
\bea
a\ket{0}=0\,,
\eea
is  a maximum energy, rather than a minimum energy, state. Also,
while the positive norm states for the $\phi$ fields are the positive
frequency states, $e^{-i\omega t}$,  the positive norm states for the $\psi$
field are negative frequency states $e^{i\omega t}$. Thus, the outgoing
positive norm $\phi$ states are linear combinations of the ingoing positive
norm $\phi_0$ states, and ingoing negative norm $\psi$ states, and the
annihilation operators of the outgoing $\phi$ field are linear combinations of
the annihilation of the ingoing $\phi$ field and creation operators of the
ingoing $\psi$ field.  This is precisely the situation examined in the first
section. 

The fact that the $\psi$ field has negative energy is clearly an approximation
in any real world situation, as the energy will not go to $-\infty$ in
reality. However, in amplifiers, the system is often set up such that some of
the modes of the system are just this type of negative energy modes at least
for small enough perturbations of the system. In a laser, for example, pumping
the atoms to their excited state (population inversion)  gives a systems where small fluctuations in
the state of the atoms  are
of exactly the above type. They can be treated as if one had a field with
negative energy. The ``ground state", the state in which all of the atoms are
in the excited state, has linear fluctuations which decrease the energy of the
system of atoms. Of course, at large
amplitudes, those modes will saturate and the system will become non-linear
(when a significant portion of the atoms have made the transition from the
excited state to the ground state). Thus this model is not a good model for
 the non-linear  regimes of such an amplifier, but is a good approximation as long as one
is concerned only with its small signal behaviour. Note that that small signal
regime can be one in which the excitations of the $\psi$ field are much much
larger than the size of the mean quantum or thermal noise in that field.

The annihilation operators for a specific mode $\phi_i(t-x)$, assumed in the
distant future to be far from the origin $x=0$ is 
\bea
a_{\phi_i}=\left<\phi_i,\Phi\right> = i\left[\int \left(\phi_i^*\Pi_{\Phi}(t,x)
-\partial_t\phi_i(t,x)^*\Phi(t,x)\right) dx\right] \,,
\eea
where $\Phi$ and $\Pi_\Phi$ are  the quantum field and conjugate momentum
operators in the Heisenberg representation. The annihilation and creation
operators
obey the usual commutation relation $[a_{\phi_i},a^\dagger_{\phi_i}]=i
\left<\phi_i,\phi_i\right>$, which, if $\phi_i$ is normalized, is the usual
commutation relation for annihilation operators. Similarly, the annihilation
operator for a $\psi$ mode $\psi_j$ 
\bea
a_{\psi_j}= \left<\phi_j,\Phi\right> = i\left[\int
\left(\psi_j^*\Pi_{\Psi}(t,x)
+\partial_t\psi_j(t,x)^*\Psi(t,x)\right) dx\right] \,.
\eea

Note that for the $\Psi$ field, the momentum is $\pi_{\Psi j} = - \partial_t
\psi_j$ because of the opposite sign in the Lagrangian.

The relation between the outgoing modes of the $\psi$ and $\phi$ fields to the
ingoing are then
\bea
a_{\phi,\omega,out}&=&  a_{\phi,\omega,in} \; \left[
{-\omega^2+\Omega^2-i\omega(\epsilon^2+\tilde\epsilon^2) \over
-\omega^2+\Omega^2-i\omega(\epsilon^2-\tilde\epsilon^2)
} \right] 
\nonumber\\ && 
+a^\dagger_{\psi,\omega,in}\; 2i \left[{ \epsilon\tilde\epsilon
\omega\over -\omega^2+\Omega^2-i\omega(\epsilon^2-\tilde\epsilon^2)
} \right]\\
&=& A_\omega \; a_{\phi,\omega,in}+B_\omega \; a^\dagger_{\psi,\omega,in}\,,
\eea
where the amplification factors obey
\bea
|A_\omega|^2-|B_\omega|^2=1\,,
\eea
as required. 

The amplification $|A_\omega|^2$is maximized 
 when  $\omega=\Omega$ 
 and is there equal to $\left({\epsilon^2+\tilde\epsilon^2\over
\epsilon^2-\tilde\epsilon^2}\right)^2$. Thus to get a large amplification  we require that
$\epsilon^2-\tilde\epsilon^2\ll\epsilon^2+\tilde\epsilon^2$.

An interesting situation occurs if we take the central oscillator to be a free
particle, so that $\Omega=0$. 
The amplification is then roughly constant for a
frequency range around $\omega=0$ until $\omega\approx
\epsilon^2-\tilde\epsilon^2$  and then falls at 6dB/octave for frequencies
higher than that,
until $\omega\approx \epsilon^2+\tilde\epsilon^2$, at which frequency the amplification goes to 1. This is just the behaviour one has for many
amplifiers -- eg the gain curve of a transistor amplifier. 

\figloc{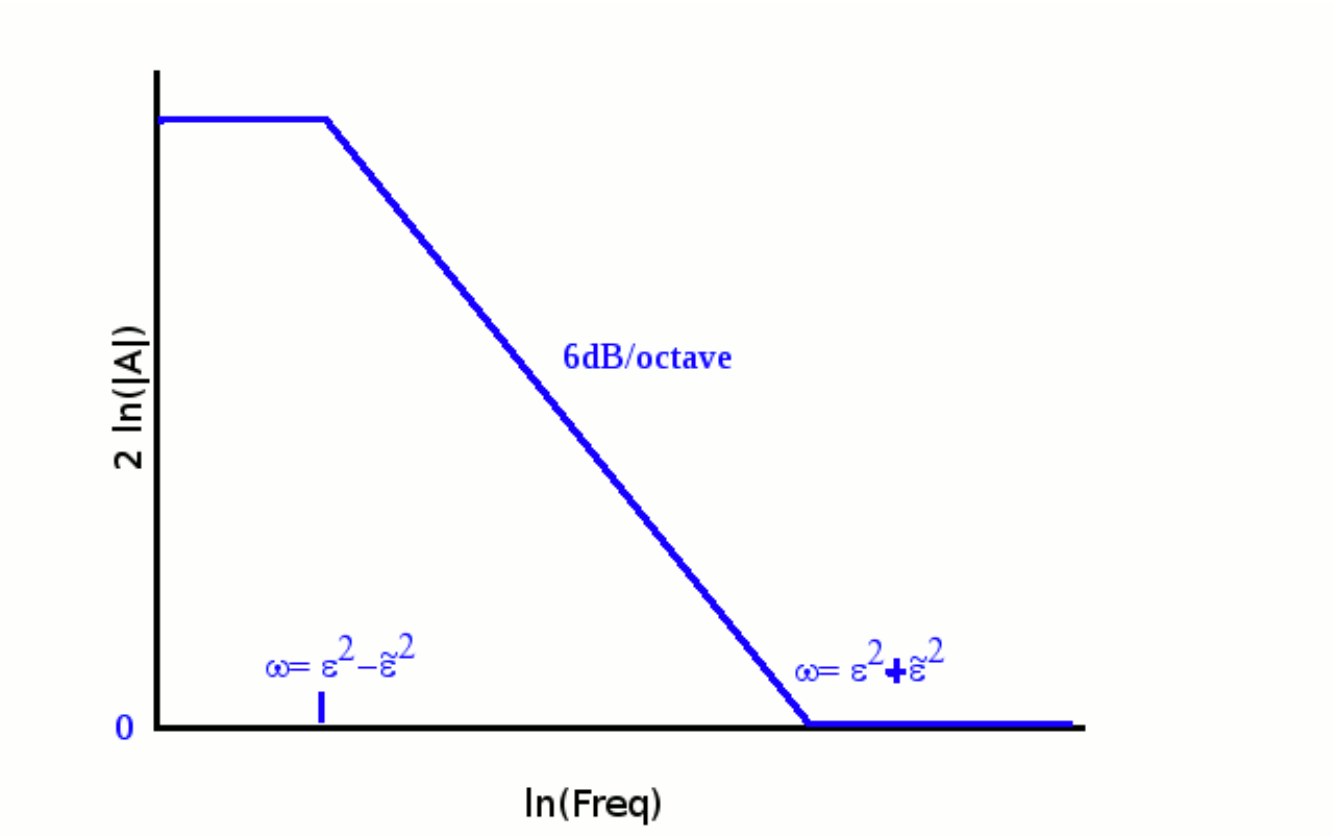}{Amplification for the $\Omega=0$ situation. Note the similarity to
the amplification of a standard integrated circuit amplifier.
}

The quantum noise temperature for the $\Omega=0$ case
 (assuming both of the input modes are at zero temperature)
and assuming $\epsilon^2\approx \tilde\epsilon^2$ is given by 
\bea
T&=&-{\omega\over \ln\left({|B|^2\over |A|^2}\right)}={\omega\over \ln\left(
{
{\omega^2\over\epsilon^4}+1}\right)}\\
&\approx& \left\{ \begin{array}{lr}{4\epsilon^4\over\omega}&\quad;
\omega \ll 2\epsilon^2\\
{\omega\over \ln\left({\omega^2\over 4\epsilon^4}\right)}&\quad;\omega \gg 2\epsilon^2
\end{array}\right\} \,.
\eea
That is, the quantum noise temperature diverges near $\omega=0$, and also for large $\omega$, and
achieves a minimum at around $\omega\approx 4\epsilon^2$. For $\omega$ smaller than that
this value, the temperature is larger than $\omega$, but for
larger $\omega$, the temperature is smaller than $\omega$ and the number of
particles in those modes will be much less than unity (ie, one is in the Wien
tail of the distribution).

Note that I have assumed that both the output and input  channels are the
$\phi$ field. One could also choose the input channel to be the $\phi$ field
and the output the $\psi$ field in which case $B_\omega$ would be the
amplification factor instead, and the amplification would fall to zero, rather
than 1 for very large frequencies. 

What is clear is that this simple model for a continuum amplifier captures
many of the features of a real amplifier and can be applied to a wide variety
of amplifiers~\cite{applications}.

\section{Black holes}

One of the more fascinating forms of amplifier is that provided by a black
hole~\cite{milburn}. Hawking~\cite{blackholes} showed that that the relationship between the ingoing and
outgoing modes of black hole could be written as
\bea
\phi_{\omega,out} = {e^{\beta\omega/2}\over \sqrt{\sinh(\beta\omega/2)}}
\; \phi^+_{\omega,in} + {e^{-\beta\omega/2}\over
\sqrt{\sinh(\beta\omega/2)}}\; \phi^-_{\omega,in} \,,
\eea
where $\phi_{\omega,out}$ is a positive norm mode escaping from the black hole with
frequency $\omega$, and $\phi^+_{\omega,in}$ is an mode ingoing toward the
star which will eventually form the black hole made up entirely of positive
norm positive frequency modes of the ingoing the field, and
$\phi^-_{\omega,in}$ are modes made up entirely of the usual negative norm,
negative frequency ingoing modes. The relation between the ingoing and
outgoing modes is unusual in that the relation between the ingoing energies
and outgoing energies is bizarre. 

Let us take as a model for black hole formation the collapse of a spherically
symmetric null shell
of dust. Before the collapse, spacetime is flat, with null coordinates $U$ and
$V$ and with ``radius" ($1\over 2\pi$ of the circumference of the spheres of
spherical symmetry) given by $r=(V-U)/2$. Let me choose the coordinate $V$ so
that $V=0$ corresponds to the shell of dust. Outside the shell the metric is
Schwarzschild, with null coordinates $u,v$ with $(v-u)/2=
(r-2M)+2M\ln({r-2M\over 2M})$, and with the null shell given by $v=0$. Along
the null shell the requirement that the circumferential radius be continuous
across the shell gives us the relation between the $U$ and $u$ coordinates as 
\bea
u=  (U+4M) -4M\ln\left(-{(U+4M)\over 4M}\right)  \,.
\eea
 Thus, if we have a wave-packet of the form of 
\bea
\phi_{out}=S(u) \; e^{i\omega u}\,,
\eea
 where $S$
is a relatively slowly varying envelope concentrated around $u \gg 4M$,
we will have that the form for the incoming wave function will be 
\bea
\phi_in(V)\approx \left\{\begin{array}{ll} 
S\left(4M{\ln(4M-V)\over 4M}\right)\left(4M-V\over 4M\right)^{-i4M\omega}
\quad\quad &;V<-4M\\
0&; V>-4M
\end{array}\right.
\eea
This can be written in terms of the ingoing positive norm modes which are pure
linear combinations of the ingoing positive norm modes $e^{-i\Omega V}$ with
$\Omega>0$.

\bea
\phi_{a,\omega}= \left\{\begin{array}{ll}
{e^{2\pi M\omega}\over\sqrt{\sinh(4\pi M\omega)}} \left({-4M-V\over
4M}\right)^{-i4M\omega} &;V<-4M\\
{e^{-2\pi M\omega}\over\sqrt{\sinh(4\pi M\omega)}} \left({4M+V\over
4M}\right)^{-i4M\omega}&;V>-4M
\end{array} \right.
\\
\phi_{b,\omega}= \left\{\begin{array}{ll}{e^{-2\pi M\omega}\over\sqrt{\sinh(4\pi M\omega)}}
\left({-4M-V\over
4M}\right)^{i4M\omega} &;V<-4M\\
{e^{2\pi M\omega}\over\sqrt{\sinh(4\pi M\omega)}} \left({4M+V\over
4M}\right)^{i4M\omega}&;V>-4M
\end{array}\right.
\eea
where in each case $\omega>0$. These two positive norm modes correspond to the
$a$, $b$ incoming modes discussed in the first section. 

Note that in this case the two types of mode $a$, $b$ correspond to different
types of the same incoming field $\phi_{in}$. The frequency of the ingoing
mode which goes as $\left({4M+V\over 4M}\right)^{-i4M\omega}$ is approximately
\bea
\Omega\approx i \partial_V \ln\left(\left({4M+V\over 4M}\right)^{-i4M\omega}\right) \approx   
            {4M\omega\over 4M+V}\approx \omega e^{u/4M}\,.
\eea
That is, the frequency of the incoming mode which creates an outgoing mode of
frequency $\omega$ at retarded time $u$ is exponential in that retarded time.
For example for a retarded time 1 second after a solar mass black hole forms,
the incoming frequency corresponding to an outgoing frequency of $\omega$ is
about $\omega e^{10^5}$ which is $e^{10^5}$ times a frequency corresponding
to the mass of the whole universe.  Thus the energy of the incoming modes
which are amplified with an amplification factor of $e^{2M\omega}\over
\sqrt{2\sinh(4\pi M\omega)}$ by the black hole amplifier, have their energy
decreased by a factor of $e^{-u\over 4M}$. Amplification does not imply energy
amplification. The black hole, as an amplifier, amplifies the amplitudes
(norms) by a thermal amplification factor, but de-amplifies the energy by a
term with is exponential in the time after the black hole forms. It is this
feature of the black hole as an amplifier that makes it unique.

\section{Dumb Holes}

In 1981 I~\cite{dumbholes} suggested that many of the features of the black hole particle
creation could also be captured by what I have since called Dumb holes~--
analogs in condensed matter system which have horizons and mimic many of the
features of black holes, including the output of quantum noise as the analog
of Hawking radiation. In the case of Hawking radiation, if one traces back the
the emitted radiation into the past, it is squeezed against the horizon
exponentially until one gets back to the time when the black hole was
originally formed. Only then can the backward propagating modes escape toward
infinity, but with absurdly high frequencies and short wavelengths. 

Dumb holes were named after the original usage of the term which  meant ``unable to
speak", and not the more modern meaning of  ``stupid"-- i.e., ``Stumm" not
``Dumm" in German, or ``Muet" not ``Stupide" in French. In analogy with black
holes which are objects which emit no light, dumb holes are objects which emit
no sound. For example, if one has the flow of water over a waterfall such that
the velocity of the water exceeds the velocity of sound in the wave somewhere
in the flow, no sound can travel upstream from beyond that point. That surface
is dumb. (For sound waves in water this would require a waterfall about 10000
km high, presenting certain experimental difficulties.) This term, by analogy
can also be applied to other waves (e.g., surface waves on water) which do not
present the same experimental difficulties. 

That horizon, which separates the region from which the waves can get to
infinity, from that region from which they cannot, is defined for low
frequency, long wavelength waves, since for most matter waves, the non-trivial
(non-linear) dispersion relation will mean that different frequencies have
different velocities, and thus different horizons. 

 Such systems with wave horizons have quantum noise 
in the same way  as black holes do -- i.e.,  they emit a thermal
spectrum of radiation of that quantized wave. That temperature is determined
by the behaviour of the flow near that horizon, just as for black holes it is
determined by the behaviour of the metric of spacetime near the horizon.

In the case of dumb holes, the same thing happens as for black holes initially (finally since we
are tracing the modes backward in time?) and the backward-in-time modes are exponentially
squeezed against the horizon. But at sufficiently short wavelength the
dispersion relation, and in particular the group velocity of the modes, changes
and the mode escapes from the horizon with very short wavelengths. Depending
on the nature of the dispersion relation the outgoing mode  can either be dragged 
in from large distances (because the group velocity is now much less than the
velocity of the fluid far from the horizon in the case where the dispersion
relation makes the group velocity small at high frequencies), or can now travel
out from inside (in the case in which the dispersion relation has a group
velocity at high frequencies much larger than the velocity of the fluid inside
the horizon). 

If we assume the flow to be stationary, the waves see a time independent
situation, and, while the wavelength can change drastically, the frequency is
conserved in the lab frame (but not in the fluid frame). In Figure 2 we have a graph of such a dispersion relation in a
still fluid, in which
the group velocity falls at high frequency. If the fluid moves with some
velocity which is smaller than the long wavelength speed  of the wave in still
water, the
dispersion relation looks as in figure 3, while if the fluid is moving with a
velocity higher than the  long wavelength speed of the wave, figure 4 gives
the dispersion relation.

\figloc{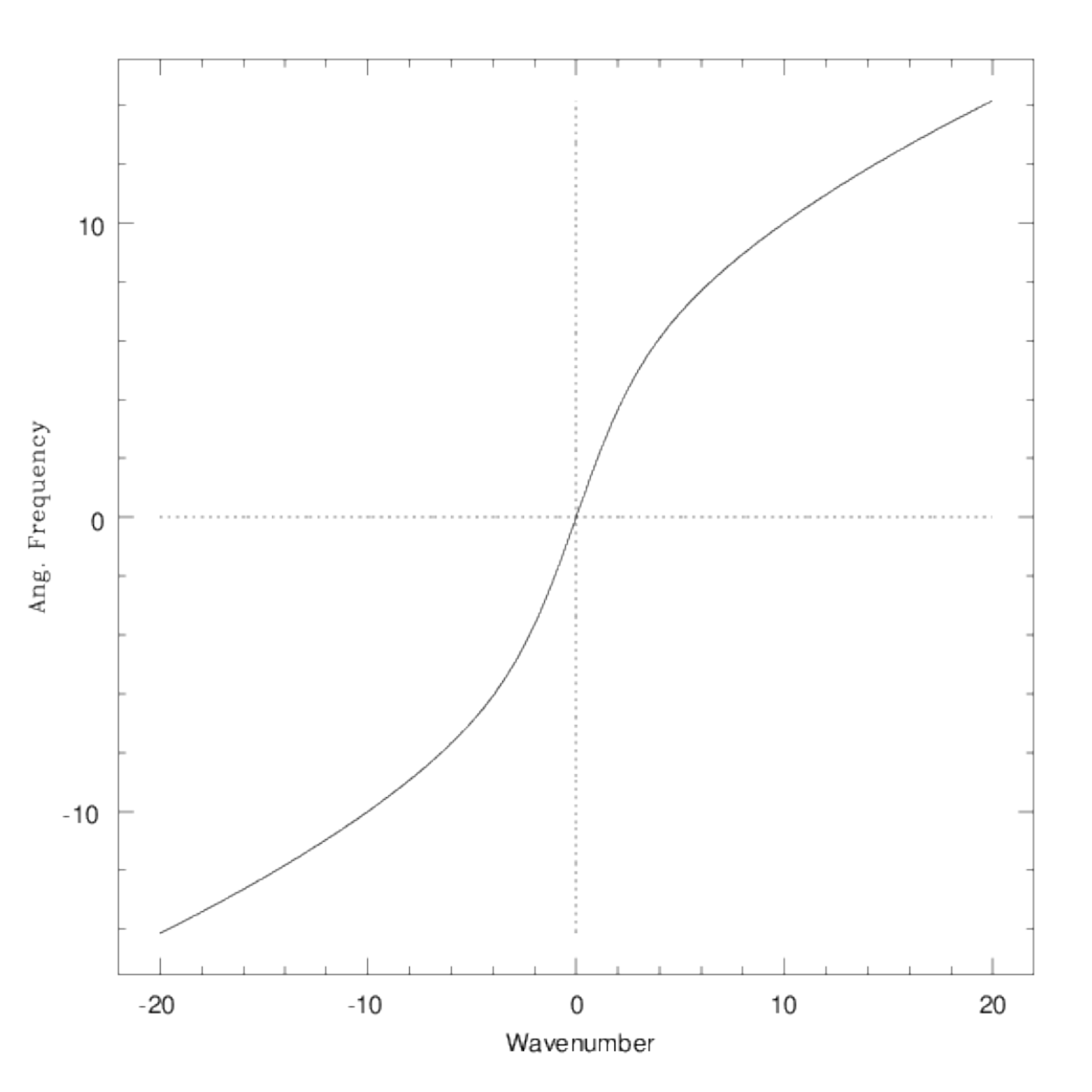}{Dispersion relation for a ``subluminal" case in still fluid. (This
is in fact the relation for surface waves on water, with
$\omega=\sqrt{gk\tanh(kh)}$).}

\figloc{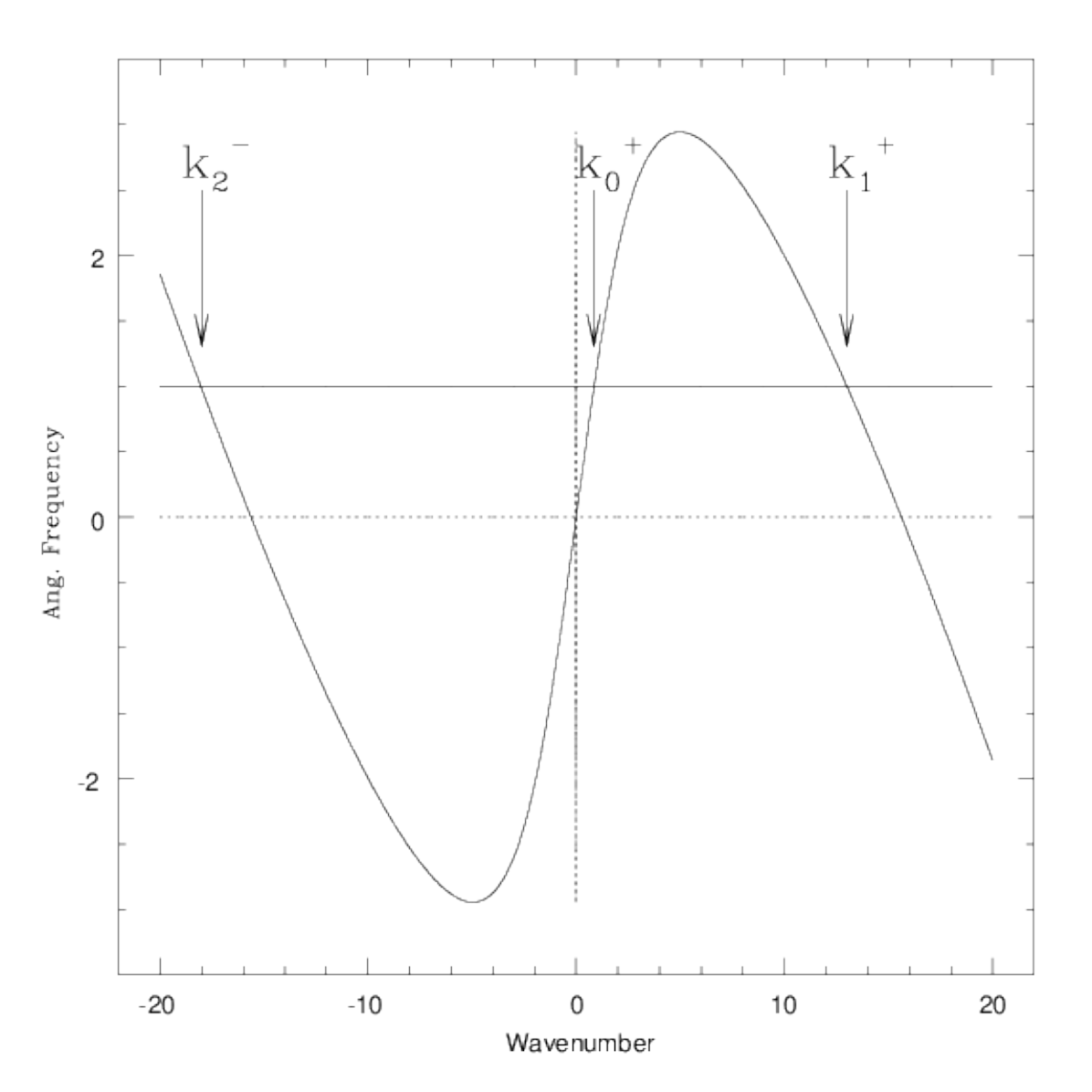}{Dispersion relation for waves as in figure 1 with the flow rate
being ``subluminal" (ie, slower than the velocity of the waves at zero
wave-vector). The horizontal line is for frequency 1, and the three possible
wave-vectors are $k_0^+$ which is a positive norm wave, both of whose phase and
group velocity is to the right. $k_1^+$ has positive phase velocity and
negative group velocity, and has positive norm. $k_2^-$ has negative phase and
group velocity and negative norm.}

\figloc{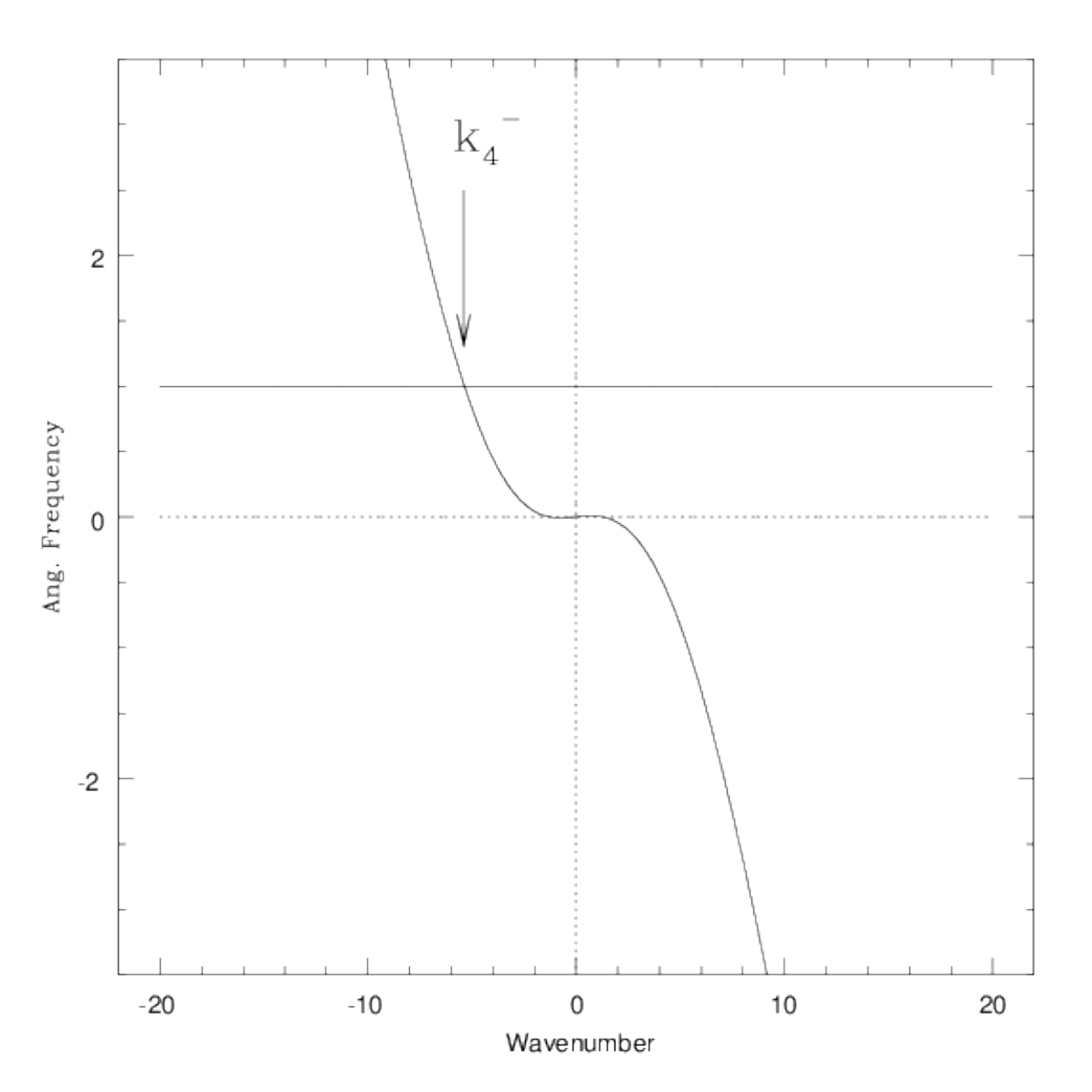}{Dispersion relation for waves as in figure 1 with the fluid flowing
faster than the velocity of the still water waves at small wave-number. 
There is only one possible wave-vector $k_4$ for the given frequency, and it has
negative phase and group velocities and negative norm.}

As an example,  let us assume we have a 1+1 dimensional wave in a fluid whose
 still fluid dispersion relation is 
\bea
\omega^2= F^2(k) \,.
\eea
In the moving fluid the dispersion relation will be 
\bea
\omega = vk \pm F(k) \,,
\eea
where the two branches represent the left and right moving waves. 
We will be interested in the minus sign which will represent waves which are
trying to move against the flow. 

The Lagrangian for such a fluid could be given by 
\bea
{\cal L}= \int (\partial_t \phi-v(x)\partial_x\phi)^2 -(F(i\partial_x) \phi)^2 \,,
\eea
where I have assumed that $F$ is an odd function of $k$.

The norm is again
\bea
\left<\phi,\phi\right>= i\int \phi^* (\partial_t\phi-v(x)\partial_x\phi)dx
+c.c.
\eea

If we assume $v$ to be a constant, and $\phi(t,x)= \phi_k e^{-i\omega t-kx} $
then the norm will be
\bea
\langle\phi,\phi\rangle \propto  2(\omega -vk)|\phi_k|^2  \,.
\eea
The important point to note is that the norm is positive or negative,
depending on the sign of $\omega-vk$. Since $\omega$ is given by the above
dispersion relation, we have 
\bea
\langle\phi,\phi\rangle = 2(F(k)) |\phi_k|^2\,.
\eea
That is, the sign of the norm depends not on the value of $\omega$ but the sign of
the still water dispersion function for that mode. Modes with positive
$\omega$ can have negative norm, and modes with negative $\omega$ can have
positive norm.

Let us imagine that we have a flow where the fast flow occurs to the right and
the slow flow to the left. There will be a horizon between the two. Now
consider modes with the frequency indicated in the diagrams. The modes which
have group velocity toward the horizon are the modes with wave-vectors $k_1$
and $k_2$. We see, that just as in the above continuum model of an amplifier,
the ingoing modes with a given positive frequency $\omega$ come in two flavours, the
ones with positive norm ($k_0$ and $\phi_0$) and ones with negative norms, but
the same positive frequencies ($k_3$ and $\psi_0$). The outgoing modes--
travelling away from the horizon, or away from the oscillator in the continuum
model-- also come in the same two flavours, the positive norm modes ($k_1$ and
$\phi_{out}$) and the negative norm outgoing modes ($k_2$ and $\psi_{out}$).
In the continuum model the coupling between these modes which leads to
amplification is the harmonic oscillator. In this dumb hole model, the
coupling between the modes is provided by the non-adiabatic, spatially
dependent changes in the background flow given by the changing velocity
$v(x)$.

In the case of the dumb hole, but not in the oscillator coupling, the
effective temperature of the emitted quantum noise from this amplifier has a
constant temperature, independent of frequency, at least for low frequencies.
This differs significantly from the continuum amplifier mentioned above where
at low frequencies the temperature
diverges~\cite{dumbholes,gravitywave,dumbholeexper}.

\section{Appendix}

To show that the $C$ state is left in a thermal density matrix after tracing
out over the $D$ states, it is easiest to do so if we assume that the state of
the system is the vacuum state for the $a$ and $b$ inputs. In this case the
condition
\bea
a\ket{0}_{ab}=b\ket{0}_{ab}
\eea
becomes
\bea
\cosh(\mu)\;C +\sinh(\mu)\;D^\dagger\ket{0}_{ab}=0\,, \\
\cosh(\mu)\;D +\sinh(\mu)\;C^\dagger\ket{0}_{ab}=0\,,
\eea
which has solution
\bea
\ket{0}_{ab}= e^{\tanh(\mu)C^\dagger D^\dagger}\ket{0}_{CD}\,,
\eea
where $\ket{0}_{CD}$ is the state annihilated by the $C$, $D$ operators.
Tracing over $D$ by using the quanta eigenstates $\ket{m}={(D^\dagger)^m\over
\sqrt{m!}}\ket{0}_{D}$
we have 
\bea
{\rm Tr}_D \ket{0}_{ab} \bra{0}_{ab} &=&\sum_m \bra{m}_D \sum_r \tanh(\mu)^r{(C^\dagger
D^\dagger)^r\over r!}\ket{0}_{CD}\nonumber\\
&& \bra{0}_{CD} \sum_s \tanh(\mu)^s {(C
D)^s\over s!}\ket{m}_D \\
&=&\sum_m  \tanh(\mu)^{2m}{(C^\dagger)^m\over \sqrt{m!}} \ket{0}_C\bra{0}_C {C^m\over
\sqrt{m!}}
\\
&=&\sum_m e^{m (2\ln(\tanh(\mu))} \ket{m}_C\bra{m}_C
\\
&=&e^{2\ln(\tanh(\mu)) C^\dagger C}\,,
\eea
which is a thermal density matrix with thermal factor~\cite{milburn}
\bea
\nonumber
x\Lambda=
-2\ln(\tanh(\mu))~\cite{milburn}
\eea

It is also clear that if one began with the two mode squeezed state
\bea
\ket{\xi}= e^{-2\ln(\tanh(\mu)) a^\dagger b^\dagger}\ket{0}_{ab}\,.
\eea
That state in terms of the output modes would just be the vacuum state
$\ket{0}_{CD}$ which would minimize the output noise.

To show that if the input state is a thermal state in each of the channels,
\bea
\rho= e^{\Lambda_a a^\dagger a} \; e^{\Lambda_b b^\dagger b} \,,
\eea
 then the output state of the $C$ channel is also a thermal state, I found it
easiest to go use path integrals.
Using 
\bea
x= {a^\dagger+a\over\sqrt{2}}\,, \hskip .5cm p_x= i\;{a^\dagger+a\over\sqrt{2}}\,,
\\
y={b^\dagger+b\over\sqrt{2}}\,, \hskip .5cm p_x= i\;{b^\dagger+b\over\sqrt{2}}\,,
\eea
we can write the density matrix,
\bea
\rho=Ne^{-{1\over 2} \Lambda_a ({\bf p_x^2+x^2}) } \,,
\eea
between the initial and final eigenstates of {\bf x,y} operators as 
\bea
\bra{x'}\rho\ket{x}&=& \int \bra{x'}\ket{p_{xN}}\bra{p_{xN}}(1-{1\over 2}
\Lambda_a ({\bf p_x^2+x^2})/N)\ket{x_{N-1}}\ldots \nonumber\\ && \ldots \bra{p_{xi}}(1-{1\over 2}
\Lambda_a ({\bf p_x^2+x^2})/N)\ket{x_{i-1}}\ldots \nonumber\\
&& \ldots \bra{p_{x1}} (1-{1\over 2}
\Lambda_a ({\bf p_x^2+x^2})/N)\ket{x}\Pi_k dp_{xk}dx_k
\\
&\approx& e^{\sum_j ip_{xj}(x_j-x_{j-1}) -\Lambda_a (p_{xj}^2 +x_j^2 ){1\over
N}}\Pi_k dp_{xk}dx_k
\\
&=& \int e^{\int_0^1 ip_x(\tau)\dot x(\tau) - {\Lambda_a\over 2}(p_x
(\tau)^2+x(\tau)^2) d\tau} \Pi_\tau dp_x(\tau)dx(\tau)\,.
\eea

Completing the squares in the exponent with respect to $p_x$ and doing the
$p_x$ integrals we get
\bea
\bra{x'}\rho\ket{x}= \int e^{-{1\over 2}\int_0^1 ({{\dot x}^2\over \Lambda}
+\Lambda x^2)d\tau }\Pi_\tau dx(\tau)\,,
\eea
where $x(1)=x'$ and $x(0)=x$. 

Similarly for the two modes, we have 
\bea
\bra{x',y'}\rho\ket{x,y}=&e^{N\int_0^1( {-{1\over 2}\int {\dot x(\tau)^2\over \Lambda_a
}+\Lambda_a x(\tau)^2 +{\dot y(\tau)^2\over \Lambda_b
}+\Lambda_b y(\tau)^2) d\tau}}\nonumber \\
&\times  \Pi_{\tau} \delta x(\tau)\delta y(\tau)\,,
\eea
where the path integral is taken over all paths $x(\tau)$, $y(\tau)$ such that
$x(0)=x$, $y(0)=y$, $x(1)=x'$, $y(1)=y'$.

As usual we can do a change of variables of the path integral, such that 
\bea
\tilde x(\tau)= x(\tau)-X(\tau)\,, \\
\tilde y(\tau)= y(\tau)-Y(\tau)\,,
\eea
where $X(\tau),~Y(\tau)$ obey
\bea
\ddot X=\Lambda_a^2 \;X\,,\\
\ddot Y=\Lambda_b^2 \;Y\,,
\eea
and where $X(0)=x,~Y(0)=y,~ X(1)=x',~Y(1)=y'$.
The boundary condition on the tilde variables is 
\bea
\tilde x(0)=\tilde x(1)=\tilde y(0)=\tilde y(1)=0\,.
\eea
The exponent of the path integral then becomes
\bea
S&=& {1\over 2}\int_0^1 {\dot x(\tau)^2\over \Lambda_a
}+\Lambda_a \; x(\tau)^2 +{\dot y(\tau)^2\over \Lambda_b
}+\Lambda_b \; y(\tau)^2)d\tau
\\
&=&{1\over 2}\int \left[{\dot X(\tau)^2\over \Lambda_a
}+\Lambda_a \; X(\tau)^2 +{\dot Y(\tau)^2\over \Lambda_b
}+\Lambda_b \; Y(\tau)^2\right] d\tau\nonumber \\ && +
{1\over 2}\int \left[{\dot {\tilde x}(\tau)^2\over \Lambda_a
}+\Lambda_a \; \tilde x(\tau)^2 +{\dot {\tilde y}(\tau)^2\over \Lambda_b
}+\Lambda_b \; \tilde y(\tau)^2\right] d\tau\,,
\eea
where the cross terms between $X$, $\tilde x$ and $Y$, $\tilde y$ vanish by integration by parts and because $X$, $Y$ obey
the equations of motion, and because $\tilde x$, $\tilde y$ are zero at the endpoints. 

Doing an integration by parts on the $X$, $Y$ terms, and using the fact that they obey
the equations of motion, we get that the only contribution to the integral is from
the endpoints. The integrand becomes
\bea
S&=&- {1\over 2} \left((x' \dot X(1) -x\dot Y(0))/\Lambda_a+(y'\dot Y(1) -y\dot
Y(0))/\Lambda_b)\right) 
\nonumber\\
&&\quad -
{1\over 2}\int \left({\dot {\tilde x}(\tau)^2\over \Lambda_a
}+\Lambda_a \tilde x(\tau)^2 +{\dot {\tilde y}(\tau)^2\over \Lambda_b
}+\Lambda_b \tilde y(\tau)^2\right)d\tau\,,
\eea
where the path integral now is over paths where the endpoints of the tilde
variables are all 0.
The contribution of the second part (the integration over the tilde variables) to the path integral is independent of the
values at the end points $x,x',y,y'$ so it simply multiplies the path integral
by a constant which can be absorbed into the normalisation factor $N$.
The solution for $X$, $Y$ with the given boundary conditions is 
\bea
X=x \;{\sinh(\Lambda_a (1-\tau))\over \sinh(\Lambda_a)}
+x' \;{\sinh(\Lambda_a\tau)\over \sinh(\Lambda_a)}\,, \\
Y=y \;{\sinh(\Lambda_b (1-\tau))\over \sinh(\Lambda_b)}
+y' \;{\sinh(\Lambda_b\tau)\over \sinh(\Lambda_b)}\,,
\eea
which gives as the only non-trivial contribution to the integrand
\bea
\rho(x'y';xy)\propto e^{\tilde S}\,,
\eea
with
\bea
\tilde S&=&-{1\over 2} \left({x' \dot X(1) -x\dot x(0)\over\Lambda_a}+{y'\dot Y(1) -y\dot
Y(0)\over\Lambda_b}\right) \nonumber \\
&=&- {1\over 2}\left((x^2+x'^2)\coth(\Lambda_a)
-2xx'{1\over\sinh(\Lambda_a)} \right. \nonumber\\ && \left. +(y^2+y'^2)\coth(\Lambda_b)
-2yy'{1\over\sinh(\Lambda_b)}\right)\,,
\eea
and 
\bea
\rho(x'y';x,y)=\bra{x'y'}\rho\ket{xy}=\tilde N e^{\tilde S}\,.
\eea
Now, we want to take the trace of this density matrix over all $D$ states.
Defining 
\bea
Z= {C+C^\dagger\over\sqrt{2}}\,, \\
W= {D+D^\dagger\over\sqrt{2}}\,,
\eea
we have that 
\bea
\ket{ x,y}= \ket{\cosh(\mu)Z-\sinh(\mu)W,\cosh(\mu)W-\sin(\mu)Z}\,,
\eea
and the trace of $\rho$ over D becomes
\bea
&&\bra{Z'}{\rm Tr}_D\rho\ket{Z}= \nonumber\\
&&\qquad\int \rho\Big({\cosh(\mu)Z'-\sinh(\mu)W,\cosh(\mu)W-\sin(\mu)Z'}; 
\nonumber \\ && 
\qquad\qquad {\cosh(\mu)Z-\sinh(\mu)W,\cosh(\mu)W-\sin(\mu)Z}\Big)\; dW\,.
\qquad
\eea

Since the integrand is an Gaussian exponential in the three variables
$Z$, $Z'$, $W$, after the integration over $W$, (since the coefficient of $W^2$ is
independent of $Z$, $Z'$) the result is also Gaussian in $Z$ and $Z'$ and is
symmetric in $Z$, $Z'$. Ie, it is also a thermal state.
Explicit calculation, by completing the squares in the exponent of the integrand
for $W$,  shows it is of the form
\bea
\bra{Z'}{\rm Tr}_D\rho\ket{Z}\propto e^{-\left({\cosh(\Lambda_C)\over
\sinh(\Lambda_C)}(Z^2+Z'^2)-{2\over\sinh(\Lambda_C)}ZZ'\right)}\,,
\eea
which is again a thermal density matrix with thermal factor $\Lambda_C$.
 While one could actually
evaluate the terms in order to determine what
$\Lambda_C$ is in terms of $\mu,~\Lambda_a,~\Lambda_b$,
it is far easier to do this by the procedure in the main section and simply
evaluate ${\rm Tr}(\rho \; C^\dagger C)$ to determine the thermal factor.

\end{document}